\documentclass[twocolumn]{aastex63}

\usepackage{soul}
\usepackage{amsmath}
\usepackage[bottom]{footmisc}
\usepackage{hyperref}
\usepackage{booktabs}

\usepackage{layouts}

\newcommand{\vlos}{v_\mathrm{los}}

\newcommand{\Rproj}{R_\mathrm{proj}}

\newcommand{\msun}{\mathrm{M}_\odot}
\newcommand{\hmsun}{h^{-1}\msun}
\newcommand{\hmpc}{h^{-1}\mathrm{Mpc}}
\newcommand{\kms}{\mathrm{km}\ \mathrm{s}^{-1}}
\newcommand{\mthc}{M_\mathrm{200c}}
\newcommand{\logmthc}{\log_{10}\left[\mthc\ \left(\hmsun\right)\right]}

\newcommand{\bfy}{\mathbf{y}}
\newcommand{\bfx}{\mathbf{x}}

\newcommand{\bfz}{\mathbf{z}}
\newcommand{\bftheta}{\boldsymbol{\theta}}
\newcommand{\bfeta}{\boldsymbol{\eta}}
\newcommand{\bfTheta}{\boldsymbol{\Theta}}
\newcommand{\bfphi}{\boldsymbol{\phi}}

\newcommand{\bbR}{\mathbb{R}}

\newcommand{\bbI}{\mathbb{I}}
\newcommand{\bbE}{\mathbb{E}}
\newcommand{\calD}{\mathcal{D}}
\newcommand{\calL}{\mathcal{L}}
\newcommand{\calF}{\mathcal{F}}
\newcommand{\calN}{\mathcal{N}}
\newcommand{\calC}{\mathcal{C}}
\newcommand{\bfg}{\mathbf{g}}

\newcommand{\msig}{$M$-$\sigma$}

\newcommand{\mtrue}{m_\mathrm{true}}

\newcommand{\Em}{\bbE\left[m|\bfx, \bfeta, \calD\right]}
\newcommand{\Vm}{\operatorname{Var}\left[m|\bfx, \bfeta, \calD\right]}

\newcommand{\point}{Point}
\newcommand{\gauss}{Gauss}
\newcommand{\class}{Class}
\newcommand{\drop}{-d}

\newcommand{\odp}{1D\point}
\newcommand{\odpd}{1D\point\drop}
\newcommand{\odg}{1D\gauss}
\newcommand{\odgd}{1D\gauss\drop}
\newcommand{\odc}{1D\class}
\newcommand{\odcd}{1D\class\drop}

\newcommand{\tdp}{2D\point}
\newcommand{\tdpd}{2D\point\drop}
\newcommand{\tdg}{2D\gauss}
\newcommand{\tdgd}{2D\gauss\drop}
\newcommand{\tdc}{2D\class}
\newcommand{\tdcd}{2D\class\drop}

\newcommand{\rev}[1]{#1}
\newcommand{\revtwo}[1]{#1}

\renewcommand{\arraystretch}{1.5}

\submitjournal{ApJ}

\shortauthors{Ho et al.}

\begin{document}
\title{Approximate Bayesian Uncertainties on Deep Learning Dynamical Mass Estimates of Galaxy Clusters}

\correspondingauthor{Matthew Ho}
\email{mho1@andrew.cmu.edu}

\author{Matthew Ho}
\affil{McWilliams Center for Cosmology,
Department of Physics, Carnegie Mellon University,
Pittsburgh, PA 15213, USA}
\affil{NSF AI Planning Institute for Physics of the Future, Carnegie Mellon University, Pittsburgh, PA 15213, USA}
\author{Arya Farahi}
\affil{The Michigan Institute for Data Science, 
University of Michigan, 
Ann Arbor, MI 48109, USA}
\author{Markus Michael Rau}
\affil{McWilliams Center for Cosmology,
Department of Physics, Carnegie Mellon University,
Pittsburgh, PA 15213, USA}
\affil{NSF AI Planning Institute for Physics of the Future, Carnegie Mellon University, Pittsburgh, PA 15213, USA}
\author{Hy Trac}
\affil{McWilliams Center for Cosmology,
Department of Physics, Carnegie Mellon University,
Pittsburgh, PA 15213, USA}
\affil{NSF AI Planning Institute for Physics of the Future, Carnegie Mellon University, Pittsburgh, PA 15213, USA}

\begin{abstract}
We study methods for reconstructing Bayesian uncertainties on dynamical mass estimates of galaxy clusters using convolutional neural networks (CNNs). We discuss the statistical background of \rev{approximate} Bayesian neural networks and demonstrate how variational inference techniques can be used to perform computationally tractable posterior estimation for a variety of deep neural architectures. We explore how various model designs and statistical assumptions impact prediction accuracy and uncertainty reconstruction in the context of cluster mass estimation. We measure the quality of our model posterior recovery using a mock cluster observation catalog derived from the MultiDark simulation and UniverseMachine catalog. We show that \rev{approximate} Bayesian CNNs produce highly accurate dynamical cluster mass posteriors. These model posteriors are log-normal in cluster mass and recover 68\% and 90\% confidence intervals to within 1\% of their measured value. We note how this rigorous modeling of dynamical mass posteriors is necessary for using cluster abundance measurements to constrain cosmological parameters.

\end{abstract}w


\section{Introduction}
Galaxy clusters are the most massive gravitationally bound systems in the universe, consisting of hundreds of luminous galaxies and hot gas embedded in dense dark matter halos. The distribution of cluster masses dominates the sensitive high mass regime of the halo mass function (HMF) and is a useful probe of large-scale structure. Measurements of cluster abundance as a function of halo mass and redshift are a major method for constraining cosmological models, but such analyses require large, well-defined cluster samples and robust mass measurement methods \citep[e.g.][]{2005RvMP...77..207V, 2011ARA&A..49..409A, 2015MNRAS.446.2205M, 2016A&A...594A..24P}. As the number of high-quality cluster observations is expected to radically increase with current and upcoming cosmological surveys such as the Dark Energy Spectroscopic Instrument (DESI), the Vera C. Rubin Observatory, and Euclid \citep{2016arXiv160407626D}, the need for precise and efficient cluster mass estimators is imperative. 

Dynamical mass estimators are a class of cluster measurements which leverage information from spectroscopic observations of member galaxies in order to infer cluster masses. The theoretical foundations of dynamical methods are grounded in the \msig\ relation, a fundamental power-law relationship which connects the mass of a stable, isotropic cluster system to the line-of-sight (LOS) velocity dispersion of its constituent galaxies. Such methods were famously used to produce the first inference of the existence of dark matter in the Coma cluster \citep{1933AcHPh...6..110Z}. Despite this historical significance, vanilla applications of the  \msig\ relation produce significant biases and scatter in realistic cluster mass predictions, owing to drastic  departures from the idealistic assumptions for which the \msig\ holds. Gravitational instabilities \citep{2018MNRAS.475..853O} and member galaxy selection effects \citep{2018MNRAS.481..324W} are prime examples of complex systematics which violate \msig\ assumptions and introduce error into dynamical cluster mass estimates. Considerable work has been done toward quantifying and mitigating the uncertainties caused by these systematics \citep[e.g.][]{2007A&A...466..437W,2013MNRAS.429.3079M,2016MNRAS.460.3900F,2018A&A...620A...8F, 2018ApJ...861...22A}. 
This proper modeling of cluster systems is crucial to the use of cluster abundance measurements for constraining cosmology.

Deep neural networks \citep[DNNs;][]{lecun2015deep} are extremely versatile machine-learning tools for modeling complex, nonlinear relationships in data-rich environments such as cosmological analyses. In recent years, DNN modeling has met a large variety of useful applications, both broadly in physics \citep[e.g.][]{2019RvMP...91d5002C} and specifically in cosmology \citep[e.g.][]{2016A&C....16...34H, 2018MNRAS.473.3895L, 2019ApJ...876...82N}. 
In \citet{2019ApJ...887...25H}, we showed that DNNs are able to mitigate systematics of dynamical cluster measurements to produce mass predictions with remarkably low bias and scatter. In addition, DNNs were computationally efficient to evaluate and robust to variations in sample richness, both requisite qualities for modern cluster mass estimators. In our comparative analysis, DNNs outperformed both simple and idealized \msig\ analyses as well as other modern machine-learning approaches \citep{2015ApJ...803...50N, 2016ApJ...831..135N, 2019MNRAS.490.2367C}

While the increasingly precise inferences produced \rev{in \citet{2019ApJ...887...25H}} prove effective for the task of point mass inference, a natural extension would be to ask how one can quantify the uncertainty of our predictions. Estimates of measurement confidence are vital to recovering Bayesian constraints on cosmological parameters. Estimating Bayesian uncertainties of deep learning models has been an exceedingly active field of study in recent years \citep[e.g.][]{neal2012bayesian, gal2016uncertainty, caldeira2020deeply}. While theoretically sound, the exact calculation of deep learning uncertainties is numerically intractable due to the necessary integration over hundreds of thousands of parameter posteriors. However, by assuming specific conjugate priors over neural network weights \citep[e.g.][]{2015arXiv150505424B, gal2016dropout}, the computational complexity of this calculation can be drastically reduced. These approximate Bayesian uncertainties have been shown to accurately recover empirical variance in a wide variety of real datasets \citep[e.g.][]{kendall2017uncertainties, 2020MNRAS.491.4277M}, with particularly strong performance in modeling out-of-sample inputs \citep[e.g.][]{gal2016dropout}.

In this paper, we seek to apply deep learning uncertainty estimation techniques to the cluster mass inference models presented in \citet{2019ApJ...887...25H}. We discuss deep learning models in a Bayesian context and how assumptions of parameter priors can be used to tractably perform weight marginalization. Using a synthetic catalog of realistic cluster observations, we measure how well deep learning models can recover confidence intervals of dynamical cluster mass estimates. We investigate how choices of predictive distribution and parameter priors impact the quality of these deep learning predictions, both for individual clusters and for cosmological analyses. This paper is organized into the following sections: in Section \ref{sec:dataset}, we describe the generation of the mock cluster catalog. In Section \ref{sec:method}, we detail the theoretical considerations for Bayesian deep learning as well as the specific designs of the presented models. In Section \ref{sec:results}, we evaluate model performance empirically and discuss the results. We summarize conclusions in Section \ref{sec:conclusion}. The code developed for this analysis is made publicly available on Github\footnote{\href{https://github.com/McWilliamsCenter/halo_cnn}{https://github.com/McWilliamsCenter/halo\_cnn}\label{foot:github}}.
\section{Dataset} \label{sec:dataset}
In this section, we summarize important properties of the mock cluster observations used in this analysis. The mock catalog is a new realization of the contaminated mock observation procedure described in \citet{2019ApJ...887...25H}. The catalog generation code is made available on Github$^\text{\ref{foot:github}}$ and pregenerated catalogs are available upon request. 

The catalog is generated from a $z=0.117$ snapshot of the MultiDark Planck 2 $N$-body simulation \citep[MDPL2;][]{2016MNRAS.457.4340K}, which assumes a $\Lambda$CDM cosmology consistent with 2013 Planck data \citep{2014A&A...571A..16P}. Host halos and subhalos are identified in the MDPL2 simulation using the ROCKSTAR halo finder \citep[MDPL2 Rockstar;][]{2013ApJ...762..109B}. We model clusters as host halos in the MDPL2 Rockstar catalog with spherical overdensity masses of $M_{200c}\geq 10^{13.5}\ \hmsun$. Galaxies are painted onto subhalos via the UniverseMachine galaxy assignment procedure \citep{2019MNRAS.488.3143B} and restricted to $M_\mathrm{stellar}\geq 10^{9.5}\ \hmsun$. Clusters and galaxies in our sample inherit mass, position, and velocity from their respective halos in the MDPL2 Rockstar and UniverseMachine catalogs. Throughout the paper, we use the shorthand $m$ to denote logarithmic spherical overdensity cluster masses,
\begin{equation} \label{eqn:massdef}
    m \equiv \logmthc.
\end{equation}

The dynamical observables reported for each mock cluster are the LOS velocities $\vlos$ and sky-projected radial positions $\Rproj$ of its selected member galaxies. For a given LOS, $\vlos$ and $\Rproj$ are calculated for all galaxies in a large neighborhood around each simulated cluster from the perspective of a $z=0$ observer. Member galaxies are then selected around each cluster in dynamical phase space $\{\vlos, \Rproj\}$ via a large cylindrical selection cut. The selection cylinder is centered at each true cluster center and oriented along the LOS, with half-length $v_\mathrm{cut}=2500\ \kms$ and radius $R_\mathrm{aperture}=1.6\ \hmpc$. \revtwo{Finally, after} the selection cut, valid mock clusters are further restricted to a richness cut of $N_\mathrm{gal}\geq 10$. \revtwo{This large, simplistic member cut ensures that the sample can be contaminated by interloping galaxies, cluster morphology, and halo environment effects, the likes of which have been shown to introduce considerable error into traditional dynamical mass estimates \citep[e.g.][]{2018MNRAS.475..853O, 2018MNRAS.481..324W}. Strong model performance under these conditions would suggest that our model could handle systematics in real observations even with very basic galaxy selection.}

Mock cluster observations are taken from multiple LOSs to augment the catalog and shape the mass distributions of the training, test, and validation sets. To mitigate biases introduced in model training, we construct the training set to have a constant number density of $dn/dm=10^{-5.2}\ h^{3} \mathrm{Mpc}^{-3} \mathrm{dex}^{-1}$ across all cluster masses $\mthc\geq\ 10^{13.5}\ \hmsun$. To achieve this evenly-distributed training set, abundant low-mass clusters are downsampled and scarce high-mass clusters are upsampled. The upsampling procedure involves taking additional projections of the same clusters from various LOSs. To avoid duplicate observations, these additional LOSs are distributed with roughly even spacing on the unit sphere. To emulate realistic measurement conditions, the test set is weighted to follow the theoretical HMF of the MDPL2 simulation and is comprised of exactly three orthogonal LOS projections per cluster.  Lastly, a validation set is created by taking a disjoint 10\% random sampling of the test set. \rev{In total, our catalog contains mock observations of ~90,000 unique host halos in the MDPL2 simulation, each observed at an average of 2.9 line-of-sights and with 37.6 member galaxy data points.}

\section{Method} \label{sec:method}
In this section, we discuss the deep learning models and uncertainty estimation techniques used to reconstruct cluster masses from member galaxy dynamics. Due to the variety of possible treatments of this problem, we seek to implement several model designs and investigate how they perform in the context of cluster mass estimation. We present a suite of twelve models, each with a different combination of input type, predictive distribution, and weight priors. 

\subsection{Input} \label{subsec:input}
The models presented in this paper infer cluster masses from one of two member galaxy distributions: the univariate distribution of LOS velocities, $\{\vlos\}$, or the joint distribution of LOS velocities and projected radial distances, $\{\vlos, \Rproj\}$. We refer to these input types as one-dimensional (1D) or two-dimensional (2D) inputs, respectively. In \citet{2019ApJ...887...25H}, we showed that the inclusion of $\Rproj$ information significantly improved the prediction performance of deep learning models. Here, we seek to investigate the impact of additional input dimensions on mass uncertainty estimation.

We use Kernel Density Estimators \citep[KDEs;][chap. 6]{scott2015multivariate} to preprocess each \revtwo{cluster's member galaxy observables} (i.e. $\vlos$ and $\Rproj$) into \revtwo{image representations} of their distributions in dynamical phase space. \revtwo{For an unknown random variable, KDEs can generate a non-parametric estimate of the probability density function (PDF) given independent samples from its distribution \citep[Eq. 2 in ][]{2019ApJ...887...25H}. In our application, we use KDEs to `smooth' each cluster's list of discrete $\vlos$ and $\Rproj$ data points into a continuous estimated PDF. The nature and scale of this smoothing is determined by a chosen kernel function which, in our case, is a Gaussian kernel with a fixed bandwidth scaling factor of $h_0=0.25$. The KDE smoothing} allows our model inputs to be more robust to fluctuations in sample richness, a desirable property for galaxy-based cluster observations. 

\begin{figure*}
    \centering
    \includegraphics{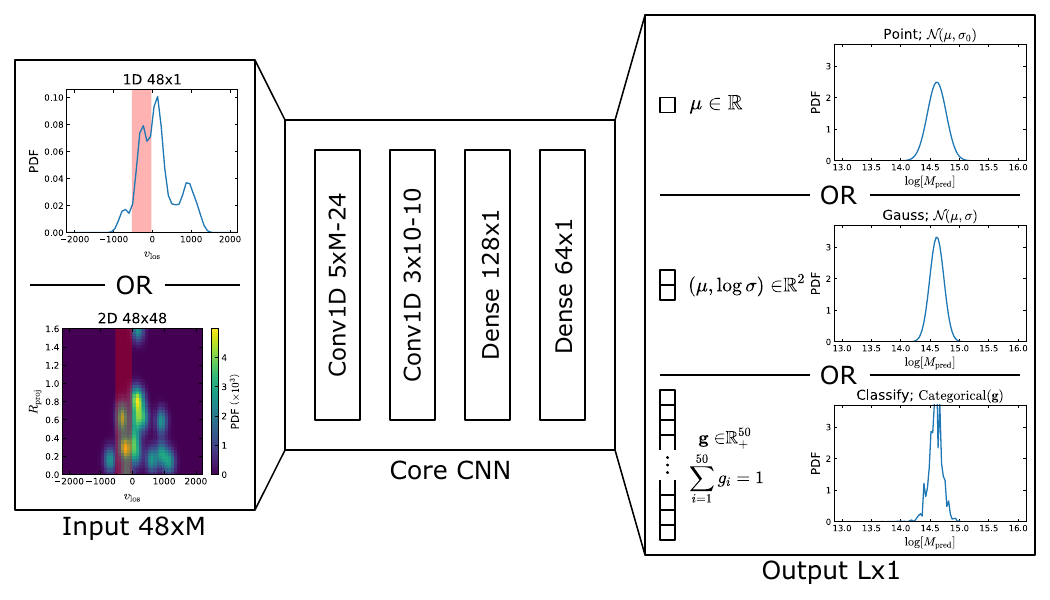}
    \caption{General convolutional neural network (CNN) architecture used for our models. In our analysis, we explore a suite of twelve models, each with different choices of inputs, outputs, and weight priors. For all models, the central core CNN architecture is identical. \rev{We generalize our input images to have shapes $48\times M$, where $M$ is equal to 1 or 48 for 1D or 2D models, respectively. All layer convolutions are taken over the $\vlos$ axis. We show an example convolutional filter highlighted in red over the input distributions.} Dropout connections are not shown here but are assumed to exist in between all layers for Dropout models. All layers utilize a rectified linear activation function (ReLU). In the diagram, convolutional layers are described using their filter shape and number of filters, respectively. Dense layers are characterized by their output layer shape. Here, we have used the notation $\mathbb{R}_+ := \{x|x\in\mathbb{R}, x>0\}$. \rev{Details of the neural architectures are further described in \S \ref{subsec:models}.}}
    \label{fig:architecture}
\end{figure*}

\revtwo{We create input images by evaluating each cluster's KDE-generated PDF at regular intervals across the dynamical phase space. 1D inputs are generated querying $\vlos$ PDFs  at 48 evenly-spaced points along the range $|\vlos| \leq v_\mathrm{cut}$. 2D inputs are derived from joint $\{\vlos,\Rproj\}$ PDFs evaluated on a regular grid of $48\times48$ points spanning the area defined by $|\vlos| \leq v_\mathrm{cut}$ and $0\leq\Rproj\leq R_\mathrm{aperture}$. Here, $v_\mathrm{cut}$ and $R_\mathrm{aperture}$ define the bounds of the cylinder cut for the mock catalog (\S\ref{sec:dataset}).} Example 1D and 2D inputs are shown in Figure \ref{fig:architecture} \revtwo{For more information on our preprocessing and a background on KDEs}, refer to \citet{2019ApJ...887...25H}.

\subsection{Deep Neural Networks}
\textit{Deep neural networks} \citep[DNNs;][]{lecun2015deep} are a class of parametric ML models which are commonly used for learning nonlinear relationships in rich, complex datasets \citep[e.g.][]{2019RvMP...91d5002C}. Within a DNN, input and output are related through a series of layered neural connections. Evaluation of a DNN involves passing input values through this sequence of neural layers, with each layer pass representing tensor multiplication with a weight matrix followed by an element-wise, nonlinear activation function. DNNs can be viewed as a functional mapping $\bfy = f\left(\bfx;\bftheta, \bfeta\right)$ between inputs $\bfx$ and outputs $\bfy$, which is parameterized by weight matrices $\bftheta$ and hyperparameters $\bfeta$ (e.g. choices of neural architecture, activation function, etc.). In general and in this application, hyperparameters $\bfeta$ are assumed to be fixed, though algorithms for optimizing these have been explored in recent literature \citep[e.g.][]{zoph2016neural}. For a more detailed explanation of DNNs and their evaluation, see \citet{2019ApJ...887...25H}.

Classically, training a DNN involves attempting to find the optimal weight parameters $\bftheta^*$ which produce the best mapping of inputs to outputs. The metric chosen to dictate model performance is called an objective loss function $\mathcal{L}$. Given a training set of example data $\calD:=\{(\bfx_i, \bfy_i)\}_{i=1}^n$, we seek to minimize this loss function over the space of possible parameters $\bfTheta$.
\begin{equation}
    \hat{\bftheta} = \arg\min_{\bftheta\in\bfTheta} \sum_{i=1}^n \mathcal{L}\left(\bfy_i, f\left(\bfx_i;\bftheta,\bfeta\right)\right)
    \label{eqn:classical}
\end{equation}
Supervised training of DNNs then becomes an optimization problem, whose solution can be found from stochastic gradient descent. Common choices of objective loss functions include mean squared error (for regression problems) and categorical cross-entropy (for classification problems). The power of neural networks arises from the fact that this optimization is numerically tractable, despite their highly nonlinear structure and thousands to millions of free parameters.

\subsection{Bayesian Uncertainties} \label{subsec:bayes}
As DNNs prove to be powerful and versatile tools for point regression and classification tasks, considerable work has gone into modeling their uncertainties \citep[e.g.][]{gal2016uncertainty}. Broadly, Bayesian uncertainties of deep learning models can be characterized as either aleatoric or epistemic. \textit{Aleatoric uncertainties} capture intrinsic scatter in input-output relationships, wherein information encoded in input data is insufficient to precisely estimate true outputs, even given an ideal model. For example, the loss of 3D dynamical information inherent in projected cluster observations introduces aleatoric uncertainties. \textit{Epistemic uncertainties} occur when training data or model flexibility is limited, such that we are unable to tightly constrain model parameters around the optimal setting, $\bftheta^*$. In the context of deep learning cluster mass estimates, epistemic uncertainties would typically arise from insufficient network depth, training time, or training catalog diversity. Specific design choices and approximations must be made for proper, computationally-tractable modeling of these uncertainties. In this paper, we investigate several of these choices in the context of deep learning cluster mass estimates. 

To capture aleatoric uncertainties, the functional output of a DNN can be used to dictate a distribution of outputs  $p(\bfy|\bfx, \bftheta, \bfeta)$ \citep[e.g.][]{1994MDN}. For example, we can train a DNN to predict parameters of a univariate Gaussian. The final layer of the network would output estimates of means and variances, $f\left(\bfx;\bftheta,\bfeta\right) = (\mu,\log \sigma)\in \bbR^2$. This framework would allow neural networks to express not only what output predictions they can make, but also the statistical confidence that they have in those predictions. The type of predictive distribution is a design choice and should be closely representative of the true conditional distribution, $p(\bfy|\bfx)$. Under ideal modeling conditions (i.e. infinite model flexibility, training data, and training time), aleatoric uncertainties are entirely sufficient for Bayesian modeling with DNNs.

However, under realistic modeling conditions, it is important to consider impacts of epistemic uncertainty on prediction. In traditional DNN training, we seek to find a single parameter setting $\hat{\bftheta}$ which optimizes some loss metric $\calL$ for the training data $\calD$. However, even with an idealized training procedure, the recovered setting $\hat{\bftheta}$ is often highly degenerate over the parameter space $\bfTheta$. When training data is limited, it is possible to recover parameter settings which minimize loss over the training set but are not representative of the data at large. To model epistemic uncertainties, we marginalize predictive distributions over the conditional probability of all possible weight parameters given the training data.
\begin{equation}
    p\left(\bfy|\bfx,\bfeta,\calD\right) = \int p\left(\bfy|\bfx,\bftheta,\bfeta\right) p\left(\bftheta|\bfeta, \calD\right) d\bftheta,
    \label{eqn:bayesian}
\end{equation}
where $p\left(\bfy|\bfx,\bfeta,\calD\right)$ is the weight-marginalized posterior distribution, $p\left(\bfy|\bfx,\bftheta,\bfeta\right)$ is the chosen predictive distribution, and $p\left(\bftheta|\bfeta, \calD\right)$ is the distribution of weight parameters informed by training data. The weight parameter distribution can be derived from Bayes rule,
\begin{equation}
    p\left(\bftheta|\bfeta, \calD\right) \propto p\left(\calD|\bftheta,\bfeta\right) p\left(\bftheta|\bfeta\right),
\end{equation}
where $p\left(\calD|\bftheta,\bfeta\right) = \prod_{i=1}^n p\left(\bfy_i|\bfx_i, \bftheta, \bfeta\right)$ and $p\left(\bftheta|\bfeta\right)$ is a chosen weight prior. Eqn. \ref{eqn:bayesian} represents exact Bayesian inference, incorporating both aleatoric and epistemic uncertainties.

\subsection{Variational Inference}

Unfortunately, the full calculation of Eqn. \ref{eqn:bayesian} is  numerically intractable for large DNNs. The integration over the space of hundreds of thousands of DNN weights is not feasible, even with highly efficient Monte Carlo methods. 
Variational inference is an alternative approach which instead interprets the posterior inference problem as an optimization. In this approach, we approximate the true weight distribution $p\left(\bftheta|\bfeta, \calD\right)$ with a  variational distribution $q(\bftheta|\hat{\bfphi})$ whose form is chosen to simplify the integration in Eqn. \ref{eqn:bayesian}. The optimal variational parameters $\hat{\bfphi}$ can then be found by minimizing the metric distance (i.e.  Kullback-Leibler divergence) between distributions $p\left(\bftheta|\bfeta, \calD\right)$ and $q(\bftheta|\hat{\bfphi})$. This minimization objective, referred to as $\calF(\calD,\bfphi)$, is often called the variational free energy or the expected lower bound (ELBO).
\begin{equation}
\begin{aligned}
    \calF(\calD,\bfphi) &= \operatorname{KL}\left[q\left(\bftheta|\bfphi\right) || p\left(\bftheta|\bfeta, \calD\right) \right]\\
    &= \bbE_{q\left(\bftheta|\bfphi\right)}\left[p\left(\calD|\bftheta,\bfeta\right)\right] 
    +\operatorname{KL}\left[q\left(\bftheta|\bfphi\right) || p\left(\bftheta|\bfeta\right) \right]
\end{aligned}
    \label{eqn:kl}
\end{equation}
where $\bbE_{q\left(\bftheta|\bfphi\right)}[\cdot]$ represents the expectation over $q\left(\bftheta|\bfphi\right)$. Equipped with the analytic forms of $q\left(\bftheta|\bfphi\right)$,  $p\left(\bfy|\bfx,\bftheta,\bfeta\right)$, and $p\left(\bftheta|\bfeta\right)$, we can minimize the objective loss in Eqn. \ref{eqn:kl} over the space of $\bfphi$'s using optimization techniques such as gradient descent. 
Under this variational technique, Bayesian posterior inference then reduces to a two-step process: a training stage wherein the optimal variational parameters $\hat{\bfphi}$ are determined from data and an inference stage which folds $q(\bftheta|\hat{\bfphi})$ into Eqn. \ref{eqn:bayesian}.

The functional forms of variational distributions $q\left(\bftheta|\bfphi\right)$ and  priors $p\left(\bftheta|\bfeta\right)$ are design choices. Several forms of variational distributions have been implemented in the literature \citep[e.g.][]{gal2016dropout, 2015arXiv150505424B}, but there lacks a consensus for an ideal choice. The most trivial variational distribution is a delta function with $q\left(\bftheta|\bfphi\right) = \delta\left(\bftheta - \bfphi\right)$. Here, we assume epistemic uncertainty to be negligible, as Eqn. \ref{eqn:bayesian} reduces to the chosen predictive distribution with $\bftheta=\hat{\bfphi}$. If we set the predictive distribution to be a fixed-mean Gaussian or a multinomial, the objective loss simplifies to the classical mean squared error or categorical cross-entropy, respectively. Furthermore, the inclusion of Gaussian or Laplacian priors $p\left(\bftheta|\bfeta\right)$, respectively, adds L2 or L1 regularization penalties to weight parameters in the loss function. 

Another common choice of weight prior is a multivariate Bernoulli distribution. \citet{gal2016dropout} investigated the nature of a Bernoulli-distributed $q\left(\bftheta|\bfphi\right)$ with a zero-mean, diagonal Gaussian $p\left(\bftheta|\bfeta\right)$. In their implementation, they utilized the popular regularization technique, Dropout, to perform stochastic integration (Eqn. \ref{eqn:bayesian}). In both the training and inference stages, Dropout layers are allowed to randomly set some fraction, $p_d\in[0,1]$, of the weight parameters equal to 0. The Dropout layers are stochastic, causing each functional evaluation of the model to use a different weight configuration. During training, this acts to regularize the iterative updates of stochastic gradient descent \citep{JMLR:v15:srivastava14a}. During inference, one can average many realizations of the Dropout layers to effectively produce a Monte Carlo estimate of the model output. \citet{gal2016dropout} showed that such a training and evaluation procedure approximates a Gaussian Process 
and is able to accurately recover uncertainties for both in- and out-of-sample data.
\newline

\begin{deluxetable*}{lCCCL}\label{tab:models}

\tablecaption{Configuration of Investigated Models}

\tablehead{\colhead{\textbf{Model Name}} & \dcolhead{\bfx} & \dcolhead{f(\bfx;\bftheta,\bfeta)} & \dcolhead{p(m|\bfx,\bftheta, \bfeta)} & \dcolhead{q(\bftheta|\bfphi)} } 

\startdata
\odp & \{\vlos\} & (\mu)\in \bbR & \calN(m;\mu,\sigma_{\calD}^2) & \prod_i\delta(\theta_i-\phi_i)\\
\odpd & \{\vlos\} & (\mu)\in \bbR & \calN(m;\mu,\sigma_{\calD}^2) & \prod_i\operatorname{Bernoulli}\left[\delta(\theta_i-\phi_i); p_d\right]\\
\odg & \{\vlos\} & (\mu,\log\sigma)\in \bbR^2 & \calN(m;\mu,\sigma^2) & \prod_i\delta(\theta_i-\phi_i)\\
\odgd & \{\vlos\} & (\mu,\log\sigma)\in \bbR^2 & \calN(m;\mu,\sigma^2) & \prod_i\operatorname{Bernoulli}\left[\delta(\theta_i-\phi_i); p_d\right]\\
\odc & \{\vlos\} & \bfg\in \bbR^{50} & \operatorname{Categorical}\left[m;S\left(\bfg\right)\right] & \prod_i\delta(\theta_i-\phi_i)\\
\odcd & \{\vlos\} & \bfg\in \bbR^{50} & \operatorname{Categorical}\left[m;S\left(\bfg\right)\right] & \prod_i\operatorname{Bernoulli}\left[\delta(\theta_i-\phi_i); p_d\right]\\\hline
\tdp & \{\Rproj, \vlos\} & (\mu)\in \bbR & \calN(m;\mu,\sigma_{\calD}^2) & \prod_i\delta(\theta_i-\phi_i)\\
\tdpd & \{\Rproj, \vlos\} & (\mu)\in \bbR & \calN(m;\mu,\sigma_{\calD}^2) & \prod_i\operatorname{Bernoulli}\left[\delta(\theta_i-\phi_i); p_d\right]\\
\tdg & \{\Rproj, \vlos\} & (\mu,\log\sigma)\in \bbR^2 & \calN(m;\mu,\sigma^2) & \prod_i\delta(\theta_i-\phi_i)\\
\tdgd & \{\Rproj, \vlos\} & (\mu,\log\sigma)\in \bbR^2 & \calN(m;\mu,\sigma^2) & \prod_i\operatorname{Bernoulli}\left[\delta(\theta_i-\phi_i); p_d\right]\\
\tdc & \{\Rproj, \vlos\} & \bfg\in \bbR^{50} & \operatorname{Categorical}\left[m;S\left(\bfg\right)\right] & \prod_i\delta(\theta_i-\phi_i)\\
\tdcd & \{\Rproj, \vlos\} & \bfg\in \bbR^{50} & \operatorname{Categorical}\left[m;S\left(\bfg\right)\right] & \prod_i\operatorname{Bernoulli}\left[\delta(\theta_i-\phi_i); p_d\right]\\
\enddata

\tablecomments{Models are presented with the design choices made for their inputs $\bfx$, functional outputs $f(\bfx;\bftheta,\bfeta)$,  predictive distributions $p(m|\bfx,\bftheta, \bfeta)$, and variational weight distribution $q(\bftheta|\bfphi)$. For \point\ models, $\sigma_{\calD}^2$ is equal to the mean squared error of model predictions after training. For clarity, the dependence of functional outputs $\mu, \sigma,$ and $\bfg$ on $(\bfx;\bftheta,\bfeta)$ has been suppressed. We use the notation $D[x;p_1,p_2,\dots]$ to denote an evaluation of the PDF of distribution $D$ with parameters $(p_1,p_2,\dots)$ at $x$. $S(\cdot)$ denotes the softmax function. 
}

\end{deluxetable*}

\subsection{Models} \label{subsec:models}

The models presented in this paper attempt to infer logarithmic cluster mass, $m$ (Eqn.~\ref{eqn:massdef}), from mappings of dynamical phase space, $\bfx$  (\S\ref{subsec:input}). All models are set up with a fixed neural architecture, $\bfeta$, and trained with a labeled set of mock cluster observations, $\calD:=\{(\bfx_i, m_i)\}_{i=1}^n$.  Using the approximate Bayesian inference techniques described in \S\ref{subsec:bayes}, each model outputs a posterior distribution over logarithmic cluster masses, $p(m|\bfx, \bfeta, \calD)$.

We investigate the impact of various design choices on the performance of our models. We implement a suite of twelve models, each with a different configuration of input type $\bfx$, predictive distribution $p(m|\bfx,\bftheta, \bfeta)$, and variational distribution $q(\bftheta|\bfphi)$. For modeling aleatoric uncertainty, we choose one of three predictive distributions: a fixed-width Gaussian (\point), a variable-width Gaussian (\gauss), and a 50-bin Categorical distribution spanning $13\leq m\leq16$ (\class). For modeling epistemic uncertainty, we implement two forms of variational distributions: a standard Dirac delta function and a Bernoulli distribution \citep{gal2016dropout}. Models implementing Dropout marginalization with a Bernoulli variational distribution are named with the suffix `-d'.   Table \ref{tab:models} contains a list of each model and its respective configuration. A schematic of each model's architecture is shown in Figure \ref{fig:architecture}. The posterior distributions and objective loss functions derived for each model are tabulated in Table \ref{tab:dist}, respectively.


\revtwo{Model architectures and other hyperparameters $\bfeta$ are the same for all models.} The core architecture of each model is a convolutional neural network \citep[CNN;][]{lecun1998gradient}. \revtwo{CNNs are widely recognized as a gold standard for solving image recognition and computer vision problems in machine learning \citep{lecun2015deep}. Their strong performance is made possible by their use of \textit{convolutional filters}, neural layers which can learn patterns in localized subregions of input data. The motivation behind our use of CNNs is based in the fact that effects which add error to dynamical mass estimates, e.g. groups of interloping galaxies, halo environments, and cluster mergers, also tend to produce artifacts in the distribution of galaxy observables. Whereas other methods attempt to directly remove or mitigate the effects of these artifacts \citep[e.g.][]{2007A&A...466..437W, 2013MNRAS.429.3079M, 2016MNRAS.460.3900F}, we intend to use CNNs to autonomously detect, account for, and even utilize these artifacts to produce improved mass estimates. While the specifics of these calculations are hidden in deep learning machinery, we show in \citet{2019ApJ...887...25H} that these CNNs evaluated on our catalog can outperform even idealized \msig\ measurements. The CNN architectures applied here are the same as those introduced in \citet{2019ApJ...887...25H} and are depicted in Figure \ref{fig:architecture}.}

\revtwo{Model weight priors $p(\bftheta|\bfeta)$ are also held constant.} For each model, we use a zero-mean, diagonal Gaussian prior on all model weights\revtwo{, i.e. $p(\bftheta|\bfeta) = \calN\left(0,\lambda \mathbb{I}\right)$ with $\lambda=10^{-4}$. This choice serves to regularize model training and avoid overfitting.} Inclusion of this prior amounts to adding a weight decay regularization term $\lambda||\bftheta||_2^2$ to each objective loss function.

\subsection{Implementation}

For models using a delta function variational distribution, training and inference are exactly equivalent to classical DNN models. Since this distribution assumes $\bftheta = \bfphi$, optimization reduces to solving Eqn. \ref{eqn:classical} via gradient descent for the loss functions shown in Table \ref{tab:dist}. Inference simplifies to an evaluation of our chosen predictive distribution at the optimized parameterization, $p\left(m|\bfx,\bfeta,\calD\right) = p\big(m|\bfx, \hat{\bftheta}, \bfeta\big)$.

We follow the procedure detailed in \citet{gal2016bayesian} to implement weight marginalization for models using Bernoulli variational distributions. During both model training and inference, we include Dropout layers after all existing neural layers in the core network architecture (Figure \ref{fig:architecture}). Dropout layers do no tensor operations, but instead randomly set some prescribed fraction of values from their input tensor equal to 0. This effectively makes the functional output of our neural network stochastic, as each pass includes random realizations of the several Dropout layers. \citet{gal2016dropout} showed that using gradient descent to minimize the loss functions in Table \ref{tab:dist} under these stochastic evaluation conditions solves Eqn. \ref{eqn:kl}. To perform inference, we approximate marginalization over the variational distribution by combining the network outputs of many realizations of the Dropout layers (Table \ref{tab:dist}). In our implementation, we set our dropout rate to $p_d=0.1$ and take $T=100$ realizations of the neural evaluation to produce inference. 

We use a 10-fold cross-validation scheme to train and evaluate our models. 
For a given fold, we train on 9/10 of the cluster
candidates in our catalog and test on the remaining,
independent 1/10. This process cycles for 10 folds until
predictions have been made for the entire test set. Cluster
candidates are grouped along with their rotated LOS duplicates
in the training-test split, such that we are never training and
testing on the same cluster from different LOSs. This ensures
independence of training and testing data for each fold. On
average, there are $\sim10,000$ training and $\sim7000$ test cluster
candidates for a given fold. 

During training, we use the Adam optimization procedure \citep{2014arXiv1412.6980K}  with a learning rate of $10^{-3}$ and a batch size of $100$. We achieve loss convergence within $40$ epochs of training. All models are implemented using the \textit{Keras}\footnote{\href{https://keras.io/}{https://keras.io/}} deep learning library with a \textit{Theano}\footnote{\href{http://deeplearning.net/software/theano/}{http://deeplearning.net/software/theano/}} backend.
\section{Results} \label{sec:results}

\newcommand{\overbar}[1]{\mkern 7mu\overline{\mkern-7mu#1\mkern-7mu}\mkern 7mu}

\begin{deluxetable*}{lCCCCCCCC}[!htb]\label{tab:stats}

\tablecaption{Descriptive Statistics of Model Performance.}

\tablehead{\colhead{\textbf{Model Name}} & \dcolhead{\tilde{\epsilon}\pm \Delta\epsilon\tablenotemark{\small a}} & \dcolhead{\sigma_\epsilon\tablenotemark{\small b}} & \dcolhead{\gamma\tablenotemark{\small b}} & \dcolhead{\kappa\tablenotemark{\small b}} & \dcolhead{\overbar{\operatorname{Var}}\left[m|\bfx, \bfeta, \calD\right]\tablenotemark{\small c}} & \dcolhead{\hat{r}(0.5)\tablenotemark{\small d}} & \colhead{16-84 EPR\tablenotemark{\small d}} & \colhead{5-95 EPR\tablenotemark{\small d}} }

\startdata
1DPoint & -0.032^{+0.168}_{-0.167} & 0.172 & 0.427 & 0.758 &0.032 & 0.425 & 0.697 & 0.919\\
1DPoint-d & -0.036^{+0.170}_{-0.156} & 0.170 & 0.469 & 0.911 &0.035 & 0.415 & 0.725 & 0.931\\
1DGauss & -0.033^{+0.173}_{-0.162} & 0.173 & 0.382 & 0.774 &0.026 & 0.425 & 0.682 & 0.891\\
1DGauss-d & -0.030^{+0.167}_{-0.157} & 0.169 & 0.497 & 1.033 &0.036 & 0.427 & 0.773 & 0.945\\
1DClass & -0.034^{+0.178}_{-0.169} & 0.178 & 0.429 & 0.873 &0.030 & 0.463 & 0.673 & 0.904\\
1DClass-d & -0.045^{+0.181}_{-0.160} & 0.178 & 0.581 & 1.203 &0.033 & 0.430 & 0.715 & 0.929\\
\hline
2DPoint & -0.024^{+0.129}_{-0.129} & 0.138 & 0.362 & 1.385 &0.024 & 0.422 & 0.752 & 0.935\\
2DPoint-d & -0.030^{+0.127}_{-0.126} & 0.134 & 0.353 & 1.372 &0.027 & 0.406 & 0.776 & 0.949\\
2DGauss & -0.011^{+0.119}_{-0.125} & 0.132 & 0.193 & 1.535 &0.018 & 0.460 & 0.708 & 0.915\\
2DGauss-d & -0.003^{+0.110}_{-0.113} & 0.123 & 0.333 & 1.886 &0.023 & 0.488 & 0.778 & 0.947\\
2DClass & -0.030^{+0.128}_{-0.131} & 0.140 & 0.289 & 1.762 &0.020 & 0.433 & 0.680 & 0.904\\
2DClass-d & -0.026^{+0.125}_{-0.127} & 0.136 & 0.340 & 2.015 &0.021 & 0.446 & 0.711 & 0.925\\
\enddata
\tablenotetext{a}{Point residual median and 16-84 percentile range (dex)}
\tablenotetext{b}{Point residual standard deviation scatter (dex), skewness, and excess kurtosis, respectively\label{cite:residual}}
\tablenotetext{c}{Average posterior variance}
\tablenotetext{d}{Empirical percentile (Eqn. \ref{eqn:rhat}) median, 16-84 range, and 5-95 range, respectively. $R_1$-$R_2$ empirical percentile ranges are equivalent to $\hat{r}(R_2\%) - \hat{r}(R_1\%)$.}
\tablecomments{Quantities are averaged over all cross-validation folds of test clusters in the mass range $14\leq\mtrue\leq15$.}

\end{deluxetable*}

We quantify the validity of our uncertainty estimation techniques in the context of astronomical and cosmological analyses. The objectives of our analysis are twofold. First, we confirm that these models accurately reproduce the point prediction performance presented in \citet{2019ApJ...887...25H}. Second, we characterize the nature of our uncertainty predictions, including how well our predictive distributions match the empirical distribution of cluster masses. 
All analyses are conducted on the contaminated cluster catalog described in \S\ref{sec:dataset} wherein true masses are known. Model predictions are made using the 10-fold training and inference procedure described in \S\ref{subsec:models}.

\subsection{Point Predictions} \label{subsec:point}

\begin{figure}
    \centering
    \includegraphics[width=\linewidth]{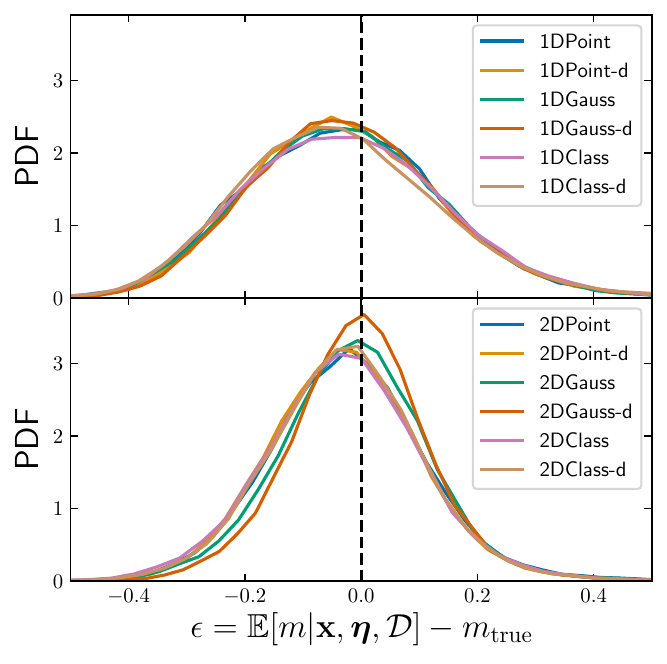}
    \caption{Distribution of point prediction residuals (Eqn. \ref{eqn:res}) for models in Table \ref{tab:models}. Point residual distributions are averaged over all cross-validation folds of test clusters in the mass range $14\leq\mtrue\leq15$.}
    \label{fig:res}
\end{figure}

\begin{figure*}[!htb]
    \centering
    \includegraphics[width=\linewidth]{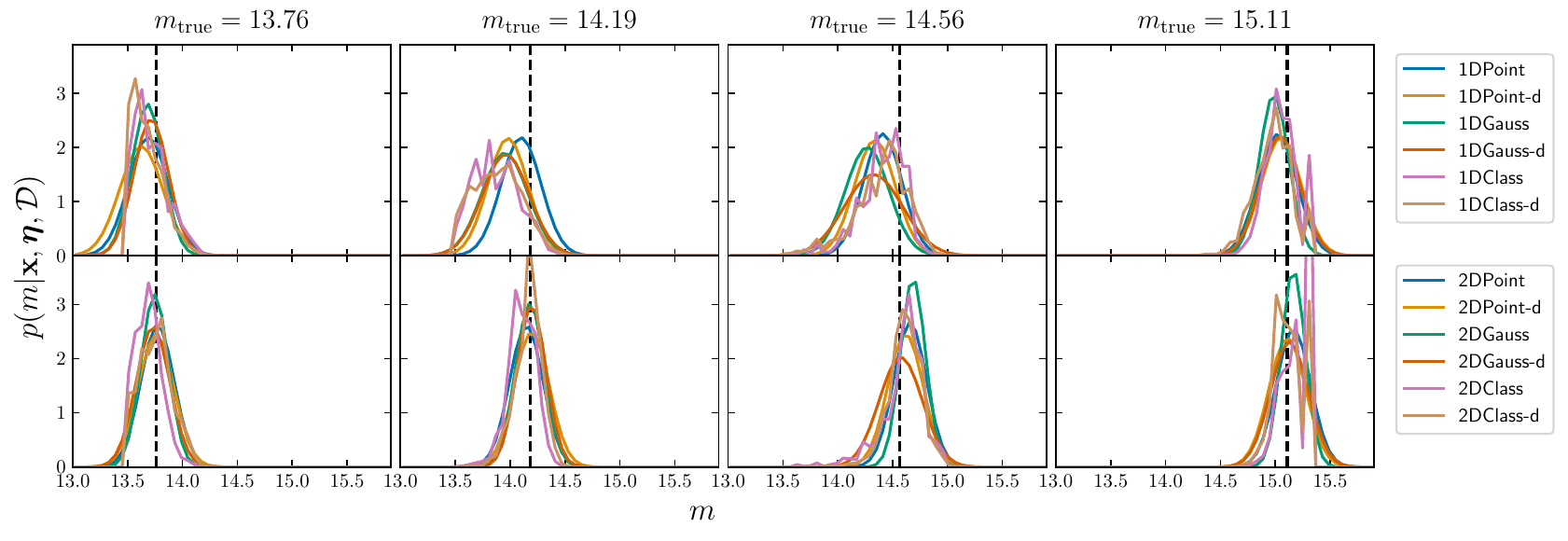}
    \caption{Posterior distributions 
    estimated by each model in Table \ref{tab:models} for four randomly-selected clusters across a variety of true masses. Each column shows mass posteriors generated from a single LOS projection of a mock cluster in our test set. Each cluster's true logarithmic mass, $\mtrue$, is stated in the column title and plotted as a black dashed line. 
    For clarity, 1D and 2D model distributions are shown on separate rows.}
    \label{fig:all_models}
\end{figure*}

We evaluate the accuracy and Gaussianity of point predictions made by our models. In this context, we define point predictions to be the mean of the estimated posterior distribution for logarithmic cluster mass (Eqn. \ref{eqn:massdef}), 
\begin{equation}
    \Em := \int m\ p(m|\bfx, \bfeta, \calD) dm.
\end{equation}
Following from this definition, we utilize the following characterization of the point residual $\epsilon$ as the difference between the point prediction and true logarithmic mass, 
\begin{equation} \label{eqn:res}
    \epsilon := \bbE\left[m|\bfx, \bfeta, \calD\right] - \mtrue.
\end{equation}
It is self-evident that, for this choice of point prediction, the models utilizing constant-variance Gaussian predictive distributions, \odp\ and \tdp, are functionally equivalent to the models presented in \citet{2019ApJ...887...25H} and should have equivalent performance. We also note that other choices of cumulative statistics such as the median or mode of the predictive distribution are also valid point predictors of cluster mass, though they are not considered here.

Our analysis shows that point residuals produced by each model in our suite have low scatter, demonstrate very low statistical bias, and are roughly Gaussian-distributed when averaged over the test dataset. This is shown in Figure \ref{fig:res} and Table \ref{tab:stats} where we have calculated the empirical distributions of point residuals and their cumulative statistics. The scatters of point estimate residuals for 1D and 2D models are approximately equal to those described in \citet{2019ApJ...887...25H}, where 1D and 2D scatters were recorded to be 0.174 dex and 0.132 dex, respectively. The difference in predictive scatter between 1D and 2D models is motivated by the inclusion of supplemental $\Rproj$ information in 2D inputs. In addition, measurements of skewness $\gamma$ and excess kurtosis $\kappa$ of the residual distributions are consistent with near-Gaussianity.


\subsection{Uncertainty Estimation}\label{subsec:unc}

\begin{figure*}
    \centering
    \includegraphics[width=\linewidth]{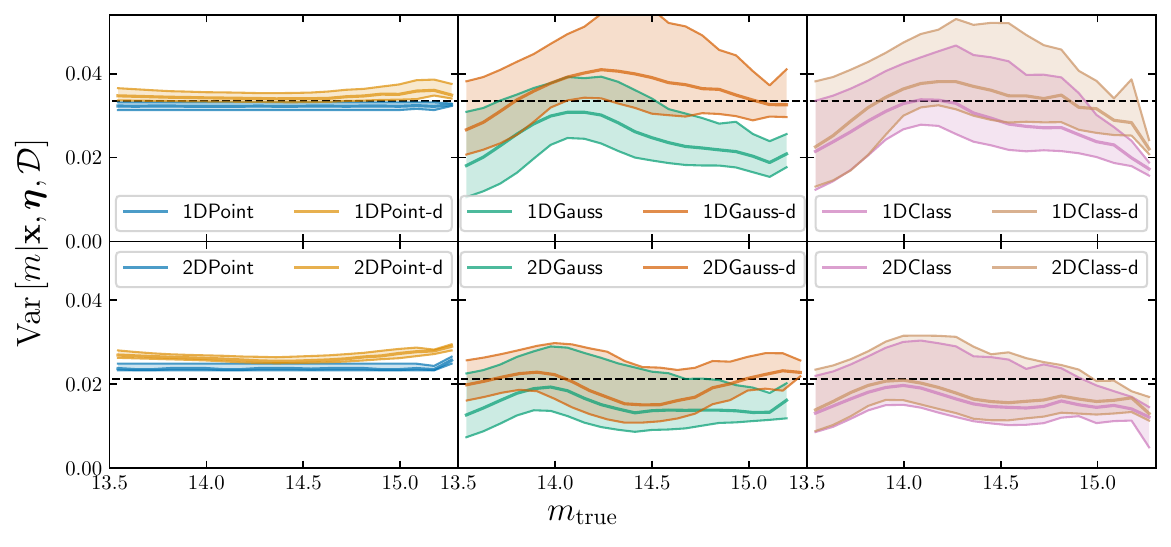}
    \caption{Posterior variance of mock cluster mass estimates (Eqn.~\ref{eqn:var}) as a function of true logarithmic mass. Standard deviation distributions are binned along true mass and shown at their median and 16-84 percentile range. Binned distributions are averaged over all cross-validation folds in the test set. For each input type, we plot a black dashed line representing the point residual variances $\sigma_\epsilon^2$, as reported by Table \ref{tab:stats}. We note that, although posterior variances of \odp\ and \tdp\ models are fixed by construction, we observe small variations in their estimates on account of the cross-validation training and evaluation procedure.}
    \label{fig:uncertainty}
\end{figure*}

\begin{figure*}
\centering
    \includegraphics[width=1\linewidth]{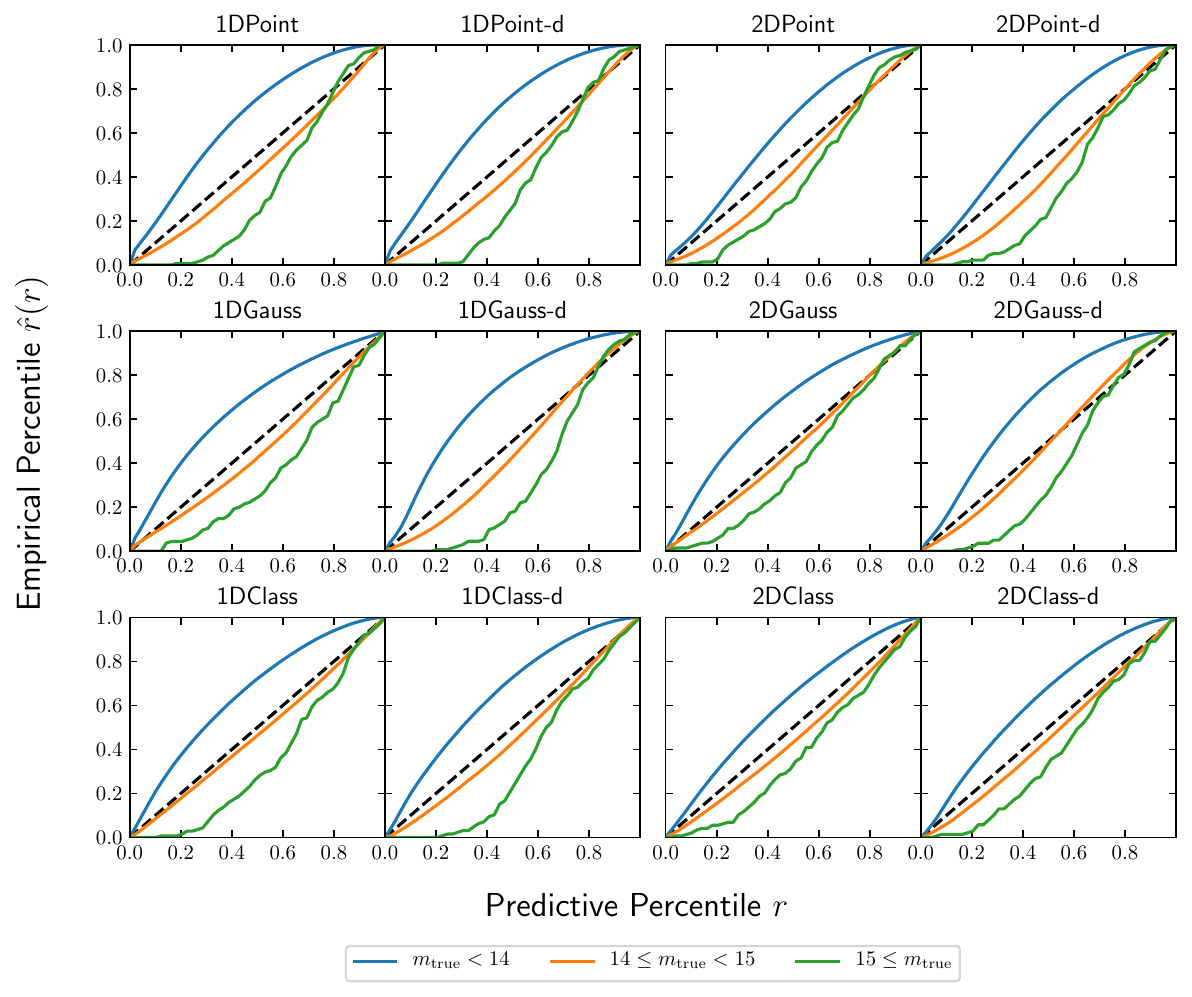}
    \caption{Empirical percentiles $\hat{r}(r)$ (Eqn. \ref{eqn:rhat}) as a function of predictive percentile $r$ for all investigated models (Table \ref{tab:models}). As implemented here, empirical percentiles capture the fraction of times the true mass of a cluster sample in our test set falls below the $r$-th quantile of our model posterior (Eqn. \ref{eqn:qtile}). We show empirical percentiles recovered for test clusters in three disjoint mass ranges.}
    \label{fig:qq}
\end{figure*}

Figure \ref{fig:all_models} shows posteriors produced by all investigated models for four randomly-selected mock clusters. In the examples shown, all models are able to accurately recover true cluster masses to within a 90\% confidence interval. Each model assigns \rev{probabilities} to small, localized regions of logarithmic masses roughly centered at $\mtrue$. Classification models only assign \rev{a non-zero probability} to mass bins where there exists training data (i.e. $13.5\leq m\leq 15.3$). For a given input type, model posteriors are strongly consistent. Classification models, whose posterior family is highly flexible, produce posteriors which are near-Gaussian, lending to the fact that assumptions of Gaussian predictive distributions for \point\ and \gauss\ models are well-founded.

To compare recovered uncertainties, we approximate each model posterior as a point estimate with Gaussian noise of variance $\Vm$. Here, we define posterior variance as:
\begin{equation}\label{eqn:var}
    \Vm := \bbE\left[m^2|\bfx, \bfeta, \calD\right] -\Em^2.
\end{equation}
The distribution of posterior variances across our test set is shown in Figure \ref{fig:uncertainty}. 

We observe that estimated posterior variances are non-constant for all models except \odp\ and \tdp, whose variances are fixed by construction. For \odpd\ and \tdpd\ models, posterior variances are largely independent of true mass. Scatter in these variance estimates arises entirely from the stochastic estimates of epistemic uncertainty. For \gauss\ and \class\ models, the estimated posterior variance exhibits a noticeable dependence on true mass. For these models, posterior variance is low for clusters at the edges of our test mass range and high for clusters around $\mtrue\sim 14$. This dependence is contrary to expectations of galaxy-based cluster mass estimators, where scatter is expected to decrease with increasing cluster richness and mass \citep{2018MNRAS.481..324W}. However, the mass-dependence of our models' posterior variances is likely biased by the mass cut placed on our training set (i.e. $\mtrue\geq13.5$). Because of this mass cut, models are trained to minimize any probability assigned to mass estimates lower than $m=13.5$. This cut thereby removes a considerable amount of variability in low mass cluster predictions. This same reasoning applies to posterior variances of high mass clusters, and its reduction effects can be observed in Figure \ref{fig:uncertainty}. However, in the safe inner range of cluster masses ($14\leq\mtrue\leq15)$, posterior variance decreases with increasing cluster mass, as expected. To mitigate the impact of the mass cut biases, future work could explore solutions such as lowering the training mass cut or reweighting low mass posteriors.

Design choices such as input type and variational distribution directly impact the magnitude of recovered posterior variances (Table \ref{tab:stats}). On average, the posterior variance estimated by 2D models is $69\%$ that of 1D models. This is expected, as the additional $\Rproj$ information given to 2D models allows recovery of tighter constraints on cluster mass \citep{2019ApJ...887...25H}. Alternatively, the use of a Bernoulli variational weight distribution over a delta function increases posterior variance by $17\%$ on average. This difference amounts to inclusion of epistemic uncertainties, which are assumed to be negligible when using a delta function.

To validate posterior recovery, we compare the posterior distributions predicted by our investigated models to the empirical distribution of true masses in the test set. To do so, we compare predictive percentiles $r$ recovered by our model posteriors to the corresponding empirical percentiles $\hat{r}(r)$ present in the data \citep{gneiting2007probabilistic}. We first define the predictive quantile $m_q\left(r;\bfx, \bfeta, \calD\right)$ as the logarithmic mass which satisfies:
\begin{equation} \label{eqn:qtile}
    r = \int_{-\infty}^{m_q\left(r;\bfx, \bfeta, \calD\right)} p\left(m|\bfx, \bfeta, \calD\right)\ dm,
\end{equation}
for a percentile $r$ and posterior $p(m|\bfx,\bftheta, \bfeta)$. We then define empirical percentile $\hat{r}(r)$ as the fraction of clusters in our test set $\calD_\mathrm{test} = \{(\bfx_i, m_i)\}_{i=1}^{N_\mathrm{test}}$ with masses less than or equal to the predictive quantile. \begin{equation}\label{eqn:rhat}
    \hat{r}(r) = \frac{1}{N_\mathrm{test}}\sum_{i=1}^{N_\mathrm{test}} \bbI \left[m_i\leq m_q\left(r;\bfx_i, \bfeta, \calD\right) \right],
\end{equation}
where $\bbI[\cdot]$ is the indicator function. Under this construction, a perfectly calibrated posterior would produce predictive percentiles which exactly match empirical percentiles, $ \hat{r}(r)=r$ for $r\in[0,1]$. A model posterior which consistently biases toward low masses would produce $\hat{r}(r)\leq r$ for $r\in[0,1]$, and vice versa for high mass biasing. If a model posterior is unbiased but underestimates variance, then $\hat{r}(r)\geq r$ for $r\in[0,0.5]$ and $\hat{r}(r)\leq r$ for $r\in[0.5,1]$, and vice versa for overestimation. This metric allows us to compare, on average, how well our models recover percentiles and confidence intervals of the true cluster mass distribution. 


Figure \ref{fig:qq} shows empirical percentiles for all investigated models. To demonstrate mass-dependent biases, we show separate lines for empirical percentiles calculated from low ($\mtrue<14$), medium ($14\leq\mtrue<15$), and high ($15\leq\mtrue$) clusters. We observe noticeable mean reversion for model predictions on the edges of our training set. All model posteriors tend to bias toward the middle of our mass range, meaning that clusters with high true masses are assigned lower mass posteriors and vice versa. This mean reversion is an inherent artifact of the interpolating behavior of machine-learning models. As non-analytic models, the DNNs implemented in this analysis struggle to extrapolate predictions to the edges of our dataset, but perform well in the inner regions. This systematic bias was also observed for point estimate masses in \citet{2016ApJ...831..135N} and \citet{2019ApJ...887...25H}. 


For observed clusters in the reliable inner mass range ($14\leq\mtrue < 15$),  predictive percentiles closely resemble empirical percentiles with a slight bias toward low mass predictions. We characterize this bias by the median empirical percentile $\hat{r}(0.5)$, as tabulated in Table \ref{tab:stats}. We find that median empirical percentiles are less than $50\%$ by at least $1.2\%$ (\tdgd) and at most $9.4\%$ (\tdpd). As a result, model predictions of median cluster mass can be expected to fall, on average, between the $40$th and $49$th percentile of the true distribution, $p(m|\bfx)$. This slight negative bias echos the findings of the point prediction analysis (\S\ref{subsec:point}) in Figure \ref{fig:res} and Table \ref{tab:stats}.

We construct a metric to quantify our models' calibration of predictive confidence intervals. We define the $R_1$-$R_2$ empirical percentile range (EPR) as the fraction of true masses captured between the $R_1$ and $R_2$th quantiles of our predictive posteriors. This quantity is equivalent to $\hat{r}(R_2\%) - \hat{r}(R_1\%)$ and should equal $R_2\%-R_1\%$ for an ideal model. Table \ref{tab:stats} tabulates the empirical percentile ranges for 16-84 and 5-95 confidence intervals. Despite median biases, model posteriors are able to recover empirical confidence intervals with a high degree of accuracy.  All models are able to recover both 16-84 and 5-95 confidence intervals to within $\pm10\%$ of their empirical value, with 2/3 of models estimating confidence intervals to within $\pm5\%$. The best performing models, \odg\ and \tdc, are able to reproduce both 16-84 and 5-95 confidence intervals to within 1\% of their empirical range. The 16-84 and 5-95 EPRs of a majority of models (2/3) tend to be greater than their fiducial values, suggesting that these models are slightly overpredicting predictive variance. 

By a small margin, \point\ models report the worst recovery of empirical percentiles among the various choices of predictive distributions. Apart from the 16-84 EPR calculated for \odp, performance metrics for \point\ models appear to deviate the most from fiducial values. The performances of \gauss\ and \class\ models appear to be roughly equal, with both model classes reporting the best recovery of empirical percentile ranges  (i.e. \odg\ and \tdc, respectively). This suggests that the added variance parameter of \gauss\ posteriors is well-utilized in our cluster mass estimation task. However, further flexibility (e.g. non-Gaussian posteriors in \class\ models) does not necessarily improve our predictive performance.

For the metrics reported in Table \ref{tab:stats}, there is little to no model performance improvement from the inclusion of a Bernoulli variational distribution. In all cases, models with Bernoulli-distributed weight priors have larger and further deviated 16-84 and 5-95 EPRs than their delta function weight prior counterparts. The inclusion of Bernoulli weight priors seems to consistently overestimate predictive uncertainties. This suggests that our training dataset is sufficiently large to tightly constrain weight parameters and that epistemic uncertainties can be safely assumed to be negligible. Other applications of Bernoulli-distributed weighting have found that improvements are very model dependent and should be tested empirically before practical application \citep{gal2016bayesian, caldeira2020deeply}.

\section{Conclusion} \label{sec:conclusion}
This paper is an extension of \citet{2019ApJ...887...25H} in which we implement modern Bayesian uncertainty reconstruction techniques for deep learning mass estimates of galaxy clusters. The deep learning models learn logarithmic cluster mass $m = \logmthc$ from dynamical cluster observables such as LOS velocities ($\vlos$) and projected radial distances ($\Rproj$) of member galaxies. We seek to estimate posterior distributions $p(m|\bfx, \bfeta, \calD)$ over cluster masses given dynamical inputs $\bfx$, network architectures $\bfeta$, and training data $\calD$. We review methods for deep learning uncertainty estimation and investigate several configurations of model design choices in our implementation. The full list of models and their respective designs is given in Table \ref{tab:models}.

We train and evaluate our models using a mock cluster observation catalog derived from a single redshift snapshot of a dark matter simulation. The mock catalog is designed to incorporate physical and selection systematics which impact real dynamical observations of galaxy clusters. We use a 10-fold cross-validation scheme to train and test our models. We measure performance metrics which characterize how well each model can both predict point estimates of cluster mass as well as recover full mass posteriors. The findings of our analysis are as follows:

\begin{itemize}
    \item To enable reconstruction of mass posteriors, we introduce additional complexity to the models first presented in \citet{2019ApJ...887...25H}. We find that this additional complexity does not diminish our ability to estimate point masses efficiently and precisely. All model implementations produce point mass estimates with Gaussian scatter at the same level as that reported in \citet{2019ApJ...887...25H}.
    \item Mass posteriors from all models in our suite are mutually consistent and assign probability to a small, localized region of cluster masses centered at the true cluster mass. The highly flexible posteriors of \class\ models converge to a near-Gaussian shape, suggesting that model assumptions of a Gaussian predictive distribution are well-founded.
    \item Inclusion of $\Rproj$ information in model inputs reduces predictive variance by $31\%$ on average. Modeling epistemic uncertainties with the Dropout approximation \citep{gal2016dropout} increases predictive variance by $17\%$ on average.
    \item Model predictions at the edges of our test set exhibit a noticeable mean-reversion effect, biasing mass posteriors toward the center of our mass range. In the inner region of our test set, model posteriors are slightly biased toward low masses on average.
    \item All models are able to recover both 16-84 and 5-95 confidence intervals to within $\pm10\%$ of their empirical value. The best performing models, \rev{\odg\ and \tdc}, are able to recover 16-84 and 5-95 confidence intervals on cluster mass to within $1\%$ of their empirical value.
    \item Modeling of epistemic uncertainties does not improve posterior recovery of our models. The impacts of epistemic uncertainties are negligible relative to posterior variances introduced by aleatoric uncertainties. This suggests that our mock catalog and training procedure are sufficient to fit the mass-observable relation.
\end{itemize}

We note that the results presented here are only tested for the simplistic mock catalogs described in \S \ref{sec:dataset} and may not necessarily hold in the presence of other realistic observational systematics such as complex survey selection functions, \revtwo{cluster miscentering}, and fiber collisions. \rev{It will be necessary to investigate the impacts of these additional systematics before this method can be applied to wide-field surveys for constraining cosmology.} We also remark that the approximate Bayesian technique described here is not the only method for reconstructing uncertainties from DNNs. An alternative method introduced by \citet{2020MNRAS.499.1985K} utilizes neural flows to infer prediction uncertainties and achieves promising results. In addition, they apply their method on spectroscopic data from the the NASA/IPAC Extragalactic Database to make preliminary dynamical mass estimates of several real galaxy clusters. 

In conclusion, we design and investigate a numerical procedure for performing approximate Bayesian inference on DNNs for galaxy cluster mass estimation. We find that this procedure is capable of recovering point estimates and confidence intervals of dynamical masses to a remarkably high degree of fidelity. The development of these uncertainty estimation techniques is a vital step toward constraining cosmology with deep learning cluster abundance measurements. Future work involving this method would investigate how more complex model inputs (e.g. 3D dynamical  phase space, multi-wavelength observations), finer tuning of hyperparameters (e.g. model architecture, KDE bandwidth), and alternative choices of variational weight distributions \citep[e.g.][]{2015arXiv150505424B} might improve recovery of mass posteriors. In addition, it will be important to study how mean-reversion biases for clusters on the low- and high-mass ends of our training catalog can be mediated in cluster abundance measurements.

We thank Fran\c{c}ois Lanusse and Michelle Ntampaka for helpful conversations and feedback while developing this project. We thank Andrew Hearin and Peter Behroozi for preparing UniverseMachine catalogs of MDPL2 simulation data. We thank John Urbanic and Doogesh Ramanah for constructive comments on our manuscript. This work is supported in part by NSF 1563887 and NSF 2020295. The computing resources necessary to complete this analysis were provided by the Pittsburgh Supercomputing Center. The CosmoSim database used in this paper is a service by the Leibniz-Institute for Astrophysics Potsdam (AIP). The MultiDark database was developed in cooperation with the Spanish MultiDark Consolider Project CSD2009-00064.
\appendix \label{sec:appendix}
This appendix contains Table \ref{tab:dist} which details explicit loss functions and output posteriors used for practical application of each investigated model.

\setcounter{table}{0}
\renewcommand{\thetable}{A\arabic{table}}

\def\arraystretch{1.5}
\begin{longrotatetable}

\begin{deluxetable}{cLL}\label{tab:dist}
\movetabledown=3em
\tablecaption{Explicit Loss Functions and Posteriors of Investigated Models}

\tablehead{\colhead{\textbf{Model Name}} & \dcolhead{\calL(\bftheta,\calD)} & \dcolhead{p\left(m|\bfx,\bfeta,\calD\right)}} 

\startdata
\point & \dfrac{1}{2n} \sum\limits_{i=1}^n \left(m_i - \mu\left(\bfx_i;\bftheta\right)\right)^2 + \lambda ||\bftheta||_2^2 & \calN\left[m; \mu(\bfx; \hat{\bftheta}), \sigma_{\calD}^2\right] \\
\point\drop & \dfrac{1}{2n} \sum\limits_{i=1}^n \left(m_i - \mu\left(\bfx_i;\bftheta\circ\bfz\right)\right)^2 + \lambda ||\bftheta||_2^2 & \calN\left[m; \hat{\bbE}_\bfz\left[\mu(\bfx; \hat{\bftheta}\circ\bfz)\right], \sigma_{\calD}^2 + \hat{\operatorname{Var}}_\bfz\left[\mu(\bfx; \hat{\bftheta}\circ\bfz)\right]\right] \\ \hline
\gauss & \dfrac{1}{2n}\sum\limits_{i=1}^n\left(\left[\left(m_i - \mu(\bfx_i;\bftheta)\right)/\sigma(\bfx_i;\bftheta)\right]^2 + \log\sigma(\bfx_i;\bftheta\circ\bfz)\right) + \lambda ||\bftheta||_2^2 & \calN\left(m; \mu(\bfx; \hat{\bftheta}), \sigma(\bfx_i;\hat{\bftheta})^2\right)\\
\gauss\drop & \dfrac{1}{2n}\sum\limits_{i=1}^n\left(\left[\left(m_i - \mu(\bfx_i;\bftheta\circ \bfz)\right)/\sigma(\bfx_i;\bftheta\circ\bfz)\right]^2 + \log\sigma(\bfx_i;\bftheta \circ \bfz)\right) + \lambda ||\bftheta||_2^2 & \calN\left(m; \hat{\bbE}_\bfz\left[\mu(\bfx; \hat{\bftheta}\circ\bfz)\right], \hat{\bbE}_\bfz\left[\sigma(\bfx; \hat{\bftheta}\circ\bfz)^2\right] + \hat{\operatorname{Var}}_\bfz\left[\mu(\bfx; \hat{\bftheta}\circ\bfz)\right]\right)\\ \hline
\class  & -\sum\limits_{i=1}^n\sum\limits_{j=1}^{50}\left(\bbI[m_i\in \calC_j]\ln g_j(\bfx_i;\bftheta) + \bbI[m_i\notin \calC_j]\ln\left[1-g_j(\bfx_i;\bftheta)\right]\right) + \lambda ||\bftheta||_2^2 & \operatorname{Categorical}\left[m;S\left(\bfg(\bfx;\hat{\bftheta})\right)\right]\\
\class\drop & -\sum\limits_{i=1}^n\sum\limits_{j=1}^{50}\left(\bbI[m_i\in \calC_j]\ln g_j(\bfx_i;\bftheta\circ\bfz) + \bbI[m_i\notin \calC_j]\ln\left[1-g_j(\bfx_i;\bftheta\circ\bfz)\right]\right) + \lambda ||\bftheta||_2^2 & \operatorname{Categorical}\left[m;S\left(\hat{\bbE}_\bfz\left[\bfg(\bfx;\hat{\bftheta}\circ\bfz)\right]\right)\right]\\
\enddata

\tablecomments{Losses and posterior forms are identical for 1D and 2D variants of the above models. For clarity, the dependence of functional outputs $\mu, \sigma$, and $\bfg$ on the fixed $\bfeta$ have been suppressed. We train and evaluate our models using a labeled training dataset, $\calD:=\{(\bfx_i, m_i)\}_{i=1}^n$. For constant variance models, $\sigma_{\calD}^2 = \frac{1}{n}\sum_{i=1}^n||m_i - \mu(\bfx_i;\hat{\bftheta})||^2_2$. To express Dropout regularization, we define the stochastic variable $\bfz := \{z_i\}_{i=1}^{\operatorname{dim}(\bftheta)}$ where $z_i\sim\operatorname{Bernoulli}(p_d)$. The operator $\circ$ represents element-wise multiplication. The operators $\hat{\bbE}_\bfz$ and $\hat{\operatorname{Var}}_\bfz$ represent empirical expectations and variances over $T=100$ i.i.d. samples of $\bfz$, respectively. $S(\cdot)$ denotes the softmax function. $C_j$ is the $j$-th bin of 50 regular classification bins spanning the mass range $13\leq m \leq 16$.}
\end{deluxetable}
\end{longrotatetable}

\newpage
\bibliography{bibliography}{}

\begin{thebibliography}{}
\expandafter\ifx\csname natexlab\endcsname\relax\def\natexlab#1{#1}\fi
\providecommand{\url}[1]{\href{#1}{#1}}
\providecommand{\dodoi}[1]{doi:~\href{http://doi.org/#1}{\nolinkurl{#1}}}
\providecommand{\doeprint}[1]{\href{http://ascl.net/#1}{\nolinkurl{http://ascl.net/#1}}}
\providecommand{\doarXiv}[1]{\href{https://arxiv.org/abs/#1}{\nolinkurl{https://arxiv.org/abs/#1}}}

\bibitem[{{Abdullah} {et~al.}(2018){Abdullah}, {Wilson}, \&
  {Klypin}}]{2018ApJ...861...22A}
{Abdullah}, M.~H., {Wilson}, G., \& {Klypin}, A. 2018, \apj, 861, 22,
  \dodoi{10.3847/1538-4357/aac5db}

\bibitem[{{Allen} {et~al.}(2011){Allen}, {Evrard}, \&
  {Mantz}}]{2011ARA&A..49..409A}
{Allen}, S.~W., {Evrard}, A.~E., \& {Mantz}, A.~B. 2011, \araa, 49, 409,
  \dodoi{10.1146/annurev-astro-081710-102514}

\bibitem[{{Behroozi} {et~al.}(2019){Behroozi}, {Wechsler}, {Hearin}, \&
  {Conroy}}]{2019MNRAS.488.3143B}
{Behroozi}, P., {Wechsler}, R.~H., {Hearin}, A.~P., \& {Conroy}, C. 2019,
  \mnras, 488, 3143, \dodoi{10.1093/mnras/stz1182}

\bibitem[{{Behroozi} {et~al.}(2013){Behroozi}, {Wechsler}, \&
  {Wu}}]{2013ApJ...762..109B}
{Behroozi}, P.~S., {Wechsler}, R.~H., \& {Wu}, H.-Y. 2013, \apj, 762, 109,
  \dodoi{10.1088/0004-637X/762/2/109}

\bibitem[{{Bishop}(1994)}]{1994MDN}
{Bishop}, M.~A. 1994, Technical Report NCRG/94/004, Aston University.
\newblock
  \url{https://publications.aston.ac.uk/id/eprint/373/1/NCRG_94_004.pdf}

\bibitem[{{Blundell} {et~al.}(2015){Blundell}, {Cornebise}, {Kavukcuoglu}, \&
  {Wierstra}}]{2015arXiv150505424B}
{Blundell}, C., {Cornebise}, J., {Kavukcuoglu}, K., \& {Wierstra}, D. 2015,
  arXiv e-prints, arXiv:1505.05424.
\newblock \doarXiv{1505.05424}

\bibitem[{Caldeira \& Nord(2020)}]{caldeira2020deeply}
Caldeira, J., \& Nord, B. 2020, Machine Learning: Science and Technology, 2,
  015002

\bibitem[{{Calderon} \& {Berlind}(2019)}]{2019MNRAS.490.2367C}
{Calderon}, V.~F., \& {Berlind}, A.~A. 2019, \mnras, 490, 2367,
  \dodoi{10.1093/mnras/stz2775}

\bibitem[{{Carleo} {et~al.}(2019){Carleo}, {Cirac}, {Cranmer}, {Daudet},
  {Schuld}, {Tishby}, {Vogt-Maranto}, \& {Zdeborov{\'a}}}]{2019RvMP...91d5002C}
{Carleo}, G., {Cirac}, I., {Cranmer}, K., {et~al.} 2019, Reviews of Modern
  Physics, 91, 045002, \dodoi{10.1103/RevModPhys.91.045002}

\bibitem[{{Dodelson} {et~al.}(2016){Dodelson}, {Heitmann}, {Hirata},
  {Honscheid}, {Roodman}, {Seljak}, {Slosar}, \&
  {Trodden}}]{2016arXiv160407626D}
{Dodelson}, S., {Heitmann}, K., {Hirata}, C., {et~al.} 2016, arXiv e-prints,
  arXiv:1604.07626.
\newblock \doarXiv{1604.07626}

\bibitem[{{Farahi} {et~al.}(2016){Farahi}, {Evrard}, {Rozo}, {Rykoff}, \&
  {Wechsler}}]{2016MNRAS.460.3900F}
{Farahi}, A., {Evrard}, A.~E., {Rozo}, E., {Rykoff}, E.~S., \& {Wechsler},
  R.~H. 2016, \mnras, 460, 3900, \dodoi{10.1093/mnras/stw1143}

\bibitem[{{Farahi} {et~al.}(2018){Farahi}, {Guglielmo}, {Evrard}, {Poggianti},
  {Adami}, {Ettori}, {Gastaldello}, {Giles}, {Maughan}, {Rapetti}, {Sereno},
  {Altieri}, {Baldry}, {Birkinshaw}, {Bolzonella}, {Bongiorno}, {Brown},
  {Chiappetti}, {Driver}, {Elyiv}, {Garilli}, {Guennou}, {Hopkins}, {Iovino},
  {Koulouridis}, {Liske}, {Maurogordato}, {Owers}, {Pacaud}, {Pierre},
  {Plionis}, {Ponman}, {Robotham}, {Sadibekova}, {Scodeggio}, {Tuffs}, \&
  {Valtchanov}}]{2018A&A...620A...8F}
{Farahi}, A., {Guglielmo}, V., {Evrard}, A.~E., {et~al.} 2018, \aap, 620, A8,
  \dodoi{10.1051/0004-6361/201731321}

\bibitem[{Gal(2016)}]{gal2016uncertainty}
Gal, Y. 2016, University of Cambridge, 1, 3.
\newblock
  \url{http://www.cs.ox.ac.uk/people/yarin.gal/website/thesis/thesis.pdf}

\bibitem[{Gal \& Ghahramani(2016{\natexlab{a}})}]{gal2016dropout}
Gal, Y., \& Ghahramani, Z. 2016{\natexlab{a}}, in International Conference on
  Machine Learning, PMLR, 1050--1059

\bibitem[{Gal \& Ghahramani(2016{\natexlab{b}})}]{gal2016bayesian}
Gal, Y., \& Ghahramani, Z. 2016{\natexlab{b}}, in 4th International Conference
  on Learning Representations (ICLR) workshop track

\bibitem[{Gneiting {et~al.}(2007)Gneiting, Balabdaoui, \&
  Raftery}]{gneiting2007probabilistic}
Gneiting, T., Balabdaoui, F., \& Raftery, A.~E. 2007, Journal of the Royal
  Statistical Society: Series B (Statistical Methodology), 69, 243

\bibitem[{{Ho} {et~al.}(2019){Ho}, {Rau}, {Ntampaka}, {Farahi}, {Trac}, \&
  {P{\'o}czos}}]{2019ApJ...887...25H}
{Ho}, M., {Rau}, M.~M., {Ntampaka}, M., {et~al.} 2019, \apj, 887, 25,
  \dodoi{10.3847/1538-4357/ab4f82}

\bibitem[{{Hoyle}(2016)}]{2016A&C....16...34H}
{Hoyle}, B. 2016, Astronomy and Computing, 16, 34,
  \dodoi{10.1016/j.ascom.2016.03.006}

\bibitem[{Kendall \& Gal(2017)}]{kendall2017uncertainties}
Kendall, A., \& Gal, Y. 2017, in Advances in neural information processing
  systems, 5574--5584

\bibitem[{{Kingma} \& {Ba}(2014)}]{2014arXiv1412.6980K}
{Kingma}, D.~P., \& {Ba}, J. 2014, arXiv e-prints, arXiv:1412.6980.
\newblock \doarXiv{1412.6980}

\bibitem[{{Klypin} {et~al.}(2016){Klypin}, {Yepes}, {Gottl{\"o}ber}, {Prada},
  \& {He{\ss}}}]{2016MNRAS.457.4340K}
{Klypin}, A., {Yepes}, G., {Gottl{\"o}ber}, S., {Prada}, F., \& {He{\ss}}, S.
  2016, \mnras, 457, 4340, \dodoi{10.1093/mnras/stw248}

\bibitem[{{Kodi Ramanah} {et~al.}(2020){Kodi Ramanah}, {Wojtak}, {Ansari},
  {Gall}, \& {Hjorth}}]{2020MNRAS.499.1985K}
{Kodi Ramanah}, D., {Wojtak}, R., {Ansari}, Z., {Gall}, C., \& {Hjorth}, J.
  2020, \mnras, 499, 1985, \dodoi{10.1093/mnras/staa2886}

\bibitem[{{Lanusse} {et~al.}(2018){Lanusse}, {Ma}, {Li}, {Collett}, {Li},
  {Ravanbakhsh}, {Mandelbaum}, \& {P{\'o}czos}}]{2018MNRAS.473.3895L}
{Lanusse}, F., {Ma}, Q., {Li}, N., {et~al.} 2018, \mnras, 473, 3895,
  \dodoi{10.1093/mnras/stx1665}

\bibitem[{LeCun {et~al.}(2015)LeCun, Bengio, \& Hinton}]{lecun2015deep}
LeCun, Y., Bengio, Y., \& Hinton, G. 2015, nature, 521, 436

\bibitem[{LeCun {et~al.}(1998)LeCun, Bottou, Bengio, \&
  Haffner}]{lecun1998gradient}
LeCun, Y., Bottou, L., Bengio, Y., \& Haffner, P. 1998, Proceedings of the
  IEEE, 86, 2278

\bibitem[{{Mamon} {et~al.}(2013){Mamon}, {Biviano}, \&
  {Bou{\'e}}}]{2013MNRAS.429.3079M}
{Mamon}, G.~A., {Biviano}, A., \& {Bou{\'e}}, G. 2013, \mnras, 429, 3079,
  \dodoi{10.1093/mnras/sts565}

\bibitem[{{Mantz} {et~al.}(2015){Mantz}, {von der Linden}, {Allen},
  {Applegate}, {Kelly}, {Morris}, {Rapetti}, {Schmidt}, {Adhikari}, {Allen},
  {Burchat}, {Burke}, {Cataneo}, {Donovan}, {Ebeling}, {Shand era}, \&
  {Wright}}]{2015MNRAS.446.2205M}
{Mantz}, A.~B., {von der Linden}, A., {Allen}, S.~W., {et~al.} 2015, \mnras,
  446, 2205, \dodoi{10.1093/mnras/stu2096}

\bibitem[{{M{\"o}ller} \& {de Boissi{\`e}re}(2020)}]{2020MNRAS.491.4277M}
{M{\"o}ller}, A., \& {de Boissi{\`e}re}, T. 2020, \mnras, 491, 4277,
  \dodoi{10.1093/mnras/stz3312}

\bibitem[{Neal(2012)}]{neal2012bayesian}
Neal, R.~M. 2012, Bayesian learning for neural networks, Vol. 118 (Springer
  Science \& Business Media)

\bibitem[{{Ntampaka} {et~al.}(2015){Ntampaka}, {Trac}, {Sutherland},
  {Battaglia}, {P{\'o}czos}, \& {Schneider}}]{2015ApJ...803...50N}
{Ntampaka}, M., {Trac}, H., {Sutherland}, D.~J., {et~al.} 2015, \apj, 803, 50,
  \dodoi{10.1088/0004-637X/803/2/50}

\bibitem[{{Ntampaka} {et~al.}(2016){Ntampaka}, {Trac}, {Sutherland},
  {Fromenteau}, {P{\'o}czos}, \& {Schneider}}]{2016ApJ...831..135N}
---. 2016, \apj, 831, 135, \dodoi{10.3847/0004-637X/831/2/135}

\bibitem[{{Ntampaka} {et~al.}(2019){Ntampaka}, {ZuHone}, {Eisenstein}, {Nagai},
  {Vikhlinin}, {Hernquist}, {Marinacci}, {Nelson}, {Pakmor}, {Pillepich},
  {Torrey}, \& {Vogelsberger}}]{2019ApJ...876...82N}
{Ntampaka}, M., {ZuHone}, J., {Eisenstein}, D., {et~al.} 2019, \apj, 876, 82,
  \dodoi{10.3847/1538-4357/ab14eb}

\bibitem[{{Old} {et~al.}(2018){Old}, {Wojtak}, {Pearce}, {Gray}, {Mamon},
  {Sif{\'o}n}, {Tempel}, {Biviano}, {Yee}, {de Carvalho}, {M{\"u}ller}, {Sepp},
  {Skibba}, {Croton}, {Bamford}, {Power}, {von der Linden}, \&
  {Saro}}]{2018MNRAS.475..853O}
{Old}, L., {Wojtak}, R., {Pearce}, F.~R., {et~al.} 2018, \mnras, 475, 853,
  \dodoi{10.1093/mnras/stx3241}

\bibitem[{{Planck Collaboration} {et~al.}(2014){Planck Collaboration}, {Ade},
  {Aghanim}, {Armitage-Caplan}, {Arnaud}, {Ashdown}, {Atrio-Barand ela},
  {Aumont}, {Baccigalupi}, {Banday}, {Barreiro}, {Bartlett}, {Battaner},
  {Benabed}, {Beno{\^\i}t}, {Benoit-L{\'e}vy}, {Bernard}, {Bersanelli},
  {Bielewicz}, {Bobin}, {Bock}, {Bonaldi}, {Bond}, {Borrill}, {Bouchet},
  {Bridges}, {Bucher}, {Burigana}, {Butler}, {Calabrese}, {Cappellini},
  {Cardoso}, {Catalano}, {Challinor}, {Chamballu}, {Chary}, {Chen}, {Chiang},
  {Chiang}, {Christensen}, {Church}, {Clements}, {Colombi}, {Colombo},
  {Couchot}, {Coulais}, {Crill}, {Curto}, {Cuttaia}, {Danese}, {Davies},
  {Davis}, {de Bernardis}, {de Rosa}, {de Zotti}, {Delabrouille}, {Delouis},
  {D{\'e}sert}, {Dickinson}, {Diego}, {Dolag}, {Dole}, {Donzelli}, {Dor{\'e}},
  {Douspis}, {Dunkley}, {Dupac}, {Efstathiou}, {Elsner}, {En{\ss}lin},
  {Eriksen}, {Finelli}, {Forni}, {Frailis}, {Fraisse}, {Franceschi}, {Gaier},
  {Galeotta}, {Galli}, {Ganga}, {Giard}, {Giardino}, {Giraud-H{\'e}raud},
  {Gjerl{\o}w}, {Gonz{\'a}lez-Nuevo}, {G{\'o}rski}, {Gratton}, {Gregorio},
  {Gruppuso}, {Gudmundsson}, {Haissinski}, {Hamann}, {Hansen}, {Hanson},
  {Harrison}, {Henrot-Versill{\'e}}, {Hern{\'a}ndez-Monteagudo}, {Herranz},
  {Hildebrand t}, {Hivon}, {Hobson}, {Holmes}, {Hornstrup}, {Hou}, {Hovest},
  {Huffenberger}, {Jaffe}, {Jaffe}, {Jewell}, {Jones}, {Juvela},
  {Keih{\"a}nen}, {Keskitalo}, {Kisner}, {Kneissl}, {Knoche}, {Knox}, {Kunz},
  {Kurki-Suonio}, {Lagache}, {L{\"a}hteenm{\"a}ki}, {Lamarre}, {Lasenby},
  {Lattanzi}, {Laureijs}, {Lawrence}, {Leach}, {Leahy}, {Leonardi},
  {Le{\'o}n-Tavares}, {Lesgourgues}, {Lewis}, {Liguori}, {Lilje},
  {Linden-V{\o}rnle}, {L{\'o}pez-Caniego}, {Lubin}, {Mac{\'\i}as-P{\'e}rez},
  {Maffei}, {Maino}, {Mand olesi}, {Maris}, {Marshall}, {Martin},
  {Mart{\'\i}nez-Gonz{\'a}lez}, {Masi}, {Massardi}, {Matarrese}, {Matthai},
  {Mazzotta}, {Meinhold}, {Melchiorri}, {Melin}, {Mendes}, {Menegoni},
  {Mennella}, {Migliaccio}, {Millea}, {Mitra}, {Miville-Desch{\^e}nes},
  {Moneti}, {Montier}, {Morgante}, {Mortlock}, {Moss}, {Munshi}, {Murphy},
  {Naselsky}, {Nati}, {Natoli}, {Netterfield}, {N{\o}rgaard-Nielsen},
  {Noviello}, {Novikov}, {Novikov}, {O'Dwyer}, {Osborne}, {Oxborrow}, {Paci},
  {Pagano}, {Pajot}, {Paladini}, {Paoletti}, {Partridge}, {Pasian},
  {Patanchon}, {Pearson}, {Pearson}, {Peiris}, {Perdereau}, {Perotto},
  {Perrotta}, {Pettorino}, {Piacentini}, {Piat}, {Pierpaoli}, {Pietrobon},
  {Plaszczynski}, {Platania}, {Pointecouteau}, {Polenta}, {Ponthieu}, {Popa},
  {Poutanen}, {Pratt}, {Pr{\'e}zeau}, {Prunet}, {Puget}, {Rachen}, {Reach},
  {Rebolo}, {Reinecke}, {Remazeilles}, {Renault}, {Ricciardi}, {Riller},
  {Ristorcelli}, {Rocha}, {Rosset}, {Roudier}, {Rowan-Robinson},
  {Rubi{\~n}o-Mart{\'\i}n}, {Rusholme}, {Sandri}, {Santos}, {Savelainen},
  {Savini}, {Scott}, {Seiffert}, {Shellard}, {Spencer}, {Starck}, {Stolyarov},
  {Stompor}, {Sudiwala}, {Sunyaev}, {Sureau}, {Sutton}, {Suur-Uski}, {Sygnet},
  {Tauber}, {Tavagnacco}, {Terenzi}, {Toffolatti}, {Tomasi}, {Tristram},
  {Tucci}, {Tuovinen}, {T{\"u}rler}, {Umana}, {Valenziano}, {Valiviita}, {Van
  Tent}, {Vielva}, {Villa}, {Vittorio}, {Wade}, {Wandelt}, {Wehus}, {White},
  {White}, {Wilkinson}, {Yvon}, {Zacchei}, \& {Zonca}}]{2014A&A...571A..16P}
{Planck Collaboration}, {Ade}, P.~A.~R., {Aghanim}, N., {et~al.} 2014, \aap,
  571, A16, \dodoi{10.1051/0004-6361/201321591}

\bibitem[{{Planck Collaboration} {et~al.}(2016){Planck Collaboration}, {Ade},
  {Aghanim}, {Arnaud}, {Ashdown}, {Aumont}, {Baccigalupi}, {Banday},
  {Barreiro}, {Bartlett}, {Bartolo}, {Battaner}, {Battye}, {Benabed},
  {Beno{\^\i}t}, {Benoit-L{\'e}vy}, {Bernard}, {Bersanelli}, {Bielewicz},
  {Bock}, {Bonaldi}, {Bonavera}, {Bond}, {Borrill}, {Bouchet}, {Bucher},
  {Burigana}, {Butler}, {Calabrese}, {Cardoso}, {Catalano}, {Challinor},
  {Chamballu}, {Chary}, {Chiang}, {Christensen}, {Church}, {Clements},
  {Colombi}, {Colombo}, {Combet}, {Comis}, {Couchot}, {Coulais}, {Crill},
  {Curto}, {Cuttaia}, {Danese}, {Davies}, {Davis}, {de Bernardis}, {de Rosa},
  {de Zotti}, {Delabrouille}, {D{\'e}sert}, {Diego}, {Dolag}, {Dole},
  {Donzelli}, {Dor{\'e}}, {Douspis}, {Ducout}, {Dupac}, {Efstathiou}, {Elsner},
  {En{\ss}lin}, {Eriksen}, {Falgarone}, {Fergusson}, {Finelli}, {Forni},
  {Frailis}, {Fraisse}, {Franceschi}, {Frejsel}, {Galeotta}, {Galli}, {Ganga},
  {Giard}, {Giraud-H{\'e}raud}, {Gjerl{\o}w}, {Gonz{\'a}lez-Nuevo},
  {G{\'o}rski}, {Gratton}, {Gregorio}, {Gruppuso}, {Gudmundsson}, {Hansen},
  {Hanson}, {Harrison}, {Henrot-Versill{\'e}}, {Hern{\'a}ndez-Monteagudo},
  {Herranz}, {Hildebrandt}, {Hivon}, {Hobson}, {Holmes}, {Hornstrup}, {Hovest},
  {Huffenberger}, {Hurier}, {Jaffe}, {Jaffe}, {Jones}, {Juvela},
  {Keih{\"a}nen}, {Keskitalo}, {Kisner}, {Kneissl}, {Knoche}, {Kunz},
  {Kurki-Suonio}, {Lagache}, {L{\"a}hteenm{\"a}ki}, {Lamarre}, {Lasenby},
  {Lattanzi}, {Lawrence}, {Leonardi}, {Lesgourgues}, {Levrier}, {Liguori},
  {Lilje}, {Linden-V{\o}rnle}, {L{\'o}pez-Caniego}, {Lubin},
  {Mac{\'\i}as-P{\'e}rez}, {Maggio}, {Maino}, {Mand olesi}, {Mangilli},
  {Maris}, {Martin}, {Mart{\'\i}nez-Gonz{\'a}lez}, {Masi}, {Matarrese},
  {McGehee}, {Meinhold}, {Melchiorri}, {Melin}, {Mendes}, {Mennella},
  {Migliaccio}, {Mitra}, {Miville-Desch{\^e}nes}, {Moneti}, {Montier},
  {Morgante}, {Mortlock}, {Moss}, {Munshi}, {Murphy}, {Naselsky}, {Nati},
  {Natoli}, {Netterfield}, {N{\o}rgaard-Nielsen}, {Noviello}, {Novikov},
  {Novikov}, {Oxborrow}, {Paci}, {Pagano}, {Pajot}, {Paoletti}, {Partridge},
  {Pasian}, {Patanchon}, {Pearson}, {Perdereau}, {Perotto}, {Perrotta},
  {Pettorino}, {Piacentini}, {Piat}, {Pierpaoli}, {Pietrobon}, {Plaszczynski},
  {Pointecouteau}, {Polenta}, {Popa}, {Pratt}, {Pr{\'e}zeau}, {Prunet},
  {Puget}, {Rachen}, {Rebolo}, {Reinecke}, {Remazeilles}, {Renault}, {Renzi},
  {Ristorcelli}, {Rocha}, {Roman}, {Rosset}, {Rossetti}, {Roudier},
  {Rubi{\~n}o-Mart{\'\i}n}, {Rusholme}, {Sandri}, {Santos}, {Savelainen},
  {Savini}, {Scott}, {Seiffert}, {Shellard}, {Spencer}, {Stolyarov}, {Stompor},
  {Sudiwala}, {Sunyaev}, {Sutton}, {Suur-Uski}, {Sygnet}, {Tauber}, {Terenzi},
  {Toffolatti}, {Tomasi}, {Tristram}, {Tucci}, {Tuovinen}, {T{\"u}rler},
  {Umana}, {Valenziano}, {Valiviita}, {Van Tent}, {Vielva}, {Villa}, {Wade},
  {Wandelt}, {Wehus}, {Weller}, {White}, {Yvon}, {Zacchei}, \&
  {Zonca}}]{2016A&A...594A..24P}
---. 2016, \aap, 594, A24, \dodoi{10.1051/0004-6361/201525833}

\bibitem[{Scott(2015)}]{scott2015multivariate}
Scott, D.~W. 2015, Multivariate density estimation: theory, practice, and
  visualization (John Wiley \& Sons)

\bibitem[{Srivastava {et~al.}(2014)Srivastava, Hinton, Krizhevsky, Sutskever,
  \& Salakhutdinov}]{JMLR:v15:srivastava14a}
Srivastava, N., Hinton, G., Krizhevsky, A., Sutskever, I., \& Salakhutdinov, R.
  2014, Journal of Machine Learning Research, 15, 1929.
\newblock \url{http://jmlr.org/papers/v15/srivastava14a.html}

\bibitem[{{Voit}(2005)}]{2005RvMP...77..207V}
{Voit}, G.~M. 2005, Reviews of Modern Physics, 77, 207,
  \dodoi{10.1103/RevModPhys.77.207}

\bibitem[{{Wojtak} {et~al.}(2007){Wojtak}, {{\L}okas}, {Mamon},
  {Gottl{\"o}ber}, {Prada}, \& {Moles}}]{2007A&A...466..437W}
{Wojtak}, R., {{\L}okas}, E.~L., {Mamon}, G.~A., {et~al.} 2007, \aap, 466, 437,
  \dodoi{10.1051/0004-6361:20066813}

\bibitem[{{Wojtak} {et~al.}(2018){Wojtak}, {Old}, {Mamon}, {Pearce}, {de
  Carvalho}, {Sif{\'o}n}, {Gray}, {Skibba}, {Croton}, {Bamford}, {Gifford},
  {von der Linden}, {Mu{\~n}oz-Cuartas}, {M{\"u}ller}, {Pearson}, {Rozo},
  {Rykoff}, {Saro}, {Sepp}, \& {Tempel}}]{2018MNRAS.481..324W}
{Wojtak}, R., {Old}, L., {Mamon}, G.~A., {et~al.} 2018, \mnras, 481, 324,
  \dodoi{10.1093/mnras/sty2257}

\bibitem[{Zoph \& Le(2016)}]{zoph2016neural}
Zoph, B., \& Le, Q.~V. 2016, arXiv preprint arXiv:1611.01578

\bibitem[{{Zwicky}(1933)}]{1933AcHPh...6..110Z}
{Zwicky}, F. 1933, Helvetica Physica Acta, 6, 110

\end{thebibliography}
\bibliographystyle{aasjournal}

\end{document}